\documentclass[sigconf]{acmart}

\AtBeginDocument{%
  \providecommand\BibTeX{{%
    \normalfont B\kern-0.5em{\scshape i\kern-0.25em b}\kern-0.8em\TeX}}}

\acmConference[IUI'24]{IUI'24: 29th International Conference on Intelligent User Interfaces}{March 18 -- 21, 2024}{Greenville, South Carolina}
\acmBooktitle{IUI'24: 29th International Conference on Intelligent User Interfaces,
  March 18 -- 21, 2024, Greenville, South Carolina}

\copyrightyear{2024}
\acmYear{2024}
\setcopyright{rightsretained}
\acmConference[IUI Companion '24]{29th International Conference on Intelligent User Interfaces - Companion}{March 18--21, 2024}{Greenville, SC, USA}
\acmBooktitle{29th International Conference on Intelligent User Interfaces - Companion (IUI Companion '24), March 18--21, 2024, Greenville, SC, USA}
\acmDOI{10.1145/3640544.3645241}
\acmISBN{979-8-4007-0509-0/24/03}

\begin{document}

\title[Charting Design Space of Human-AI Interactions]{Advancing GUI for Generative AI: Charting the Design Space of Human-AI Interactions through Task Creativity and Complexity}

\author{Zijian Ding}
\email{ding@umd.edu}
\orcid{0000-0003-4368-0336}
\affiliation{%
  \institution{University of Maryland, College Park}
  \country{USA}
}

\begin{abstract}
Technological progress has persistently shaped the dynamics of human-machine interactions in task execution. In response to the advancements in Generative AI, this paper outlines a detailed study plan that investigates various human-AI interaction modalities across a range of tasks, characterized by differing levels of creativity and complexity. This exploration aims to inform and contribute to the development of Graphical User Interfaces (GUIs) that effectively integrate with and enhance the capabilities of Generative AI systems. The study comprises three parts: exploring fixed-scope tasks through news headline generation, delving into atomic creative tasks with analogy generation, and investigating complex tasks via data visualization. Future work aims to extend this exploration to linearize complex data analysis results into narratives understandable to a broader audience, thereby enhancing the interpretability of AI-generated content.
\end{abstract}

\newcommand{\todo}[1]{{\color{blue} #1}}

\begin{teaserfigure}
\centering \includegraphics[width=0.8\textwidth]{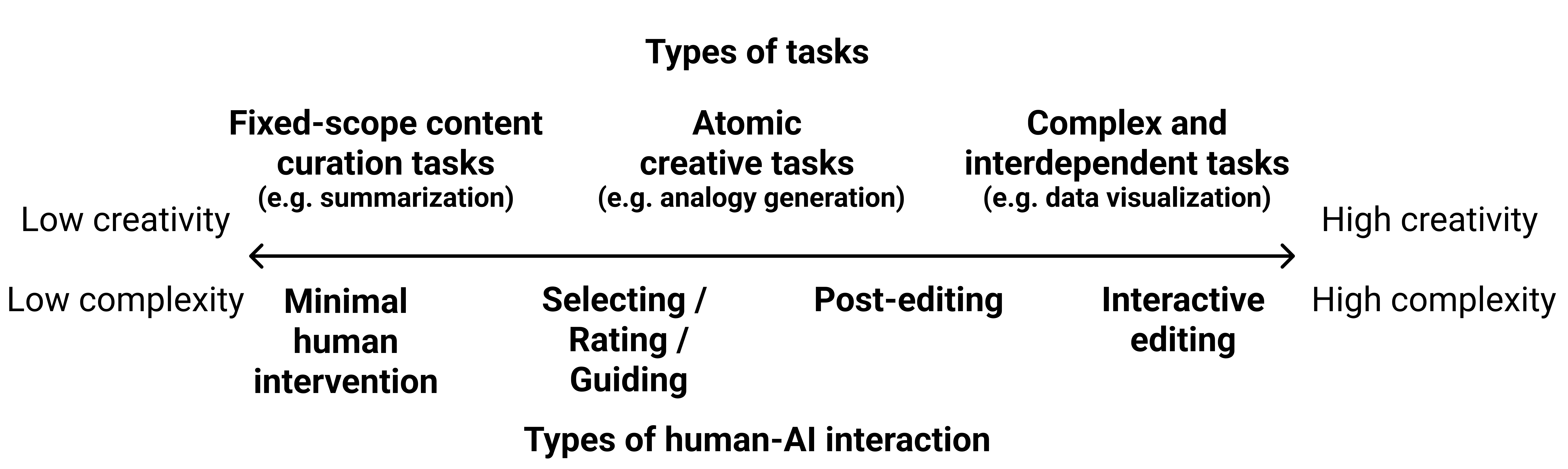}
\caption{Spectrum of human-AI co-creation tasks and corresponding human intervention complexity \cite{dingMappingDesignSpace2023}. The upper half describes the spectrum of co-creation tasks from low to high creativity and complexity; the bottom half proposes a mapping of the points on these spectrum to human-AI interaction patterns from the taxonomy proposed by Cheng et al. \cite{chengMappingDesignSpace2022}.}
\label{fig:spectrum}
\Description{TODO}
\end{teaserfigure}

\maketitle

\section{Introduction}

Over time, advancements in technology have consistently changed how humans interact with machines. The trajectory of User Interface (UI) development has experienced several pivotal shifts, starting with the inception of the Command Line Interface (CLI) in 1964, and the subsequent introduction of the Graphical User Interface (GUI) in 1979. By 2022, elements reminiscent of the CLI had resurfaced in UI design, a trend epitomized by conversational agents such as ChatGPT. The current imperative is to devise an advanced GUI iteration that seamlessly integrates with Generative AI applications. In pursuit of this, our study delves into various human-AI interaction modalities across a spectrum of task creativity and complexity to guide future GUI enhancements.

Our research categorizes tasks into three escalating levels of creativity and complexity: fixed-scope content curation tasks like summarization, atomic creative tasks like analogy generation, and complex and interdependent tasks like data analysis visualization. Correspondingly, we explore a range of human-AI interaction methods, from straightforward selection and rating to more involved guiding, post-editing, and interactive editing. This leads us to our research question: \textit{How do different facets of human-AI interaction correlate with the creative and complex nature of tasks?}

\section{Related work}

Our examination of human-AI collaboration patterns is informed by Cheng et al.'s taxonomy, which delineates five prevalent interactions in text summarization: \textit{guiding model output}, \textit{selecting or rating model output}, \textit{post-editing}, \textit{interactive editing} initiated by AI, and \textit{writing with model assistance} initiated by humans \cite{chengMappingDesignSpace2022}. My doctoral research dissects these interactions across a spectrum of tasks, varying from simple to complex and creative endeavors. This spectrum encompasses fixed-scope content curation tasks, atomic creative tasks, and complex and interdependent tasks (refer to Figure \ref{fig:spectrum}).

\subsection{Fixed-scope Content Curation Tasks}

Substantial research demonstrates that Large Language Models (LLMs) adeptly manage defined content curation tasks such as text summarization \cite{goyalNewsSummarizationEvaluation2022}, content refinement \cite{linWhyHowEmbrace2023}, and code explanation \cite{macneilGeneratingDiverseCode2022}. These tasks involve reiterating existing knowledge succinctly and coherently without generating novel insights. GPT and similar LLMs have shown proficiency in these tasks, often without the necessity for human intervention. Clark et al. revealed that text generated by GPT-3 was linguistically advanced to the extent that distinguishing it from human-written text became challenging for evaluators \cite{clarkAllThatHuman2021}. This calls for well-structured frameworks to dissect human and machine errors and ascertain authorship \cite{douGPT3TextIndistinguishable2022}. Such evidence underscores the potential for reducing human involvement in content curation tasks due to the superior quality of machine-generated text.

\subsection{Atomic Creative Tasks}

The second category comprises atomic creative tasks that necessitate generating novel and valuable outputs \cite{sawyerExplainingCreativityScience2012,chanFormulatingFixatingEffects2024}.
These include tasks such as crafting analogous design concepts \cite{zhuGenerativePreTrainedTransformer2022, leeEvaluatingHumanLanguageModel2022}, design problems \cite{macneilProbMapAutomaticallyConstructing2021,macneilFindingPlaceDesign2021,macneilFramingCreativeWork2021}, slogans \cite{clarkCreativeWritingMachine2018}, and tweetorials \cite{geroSparksInspirationScience2022}. LLMs like GPT can foster novel connections or "creative leaps" due to their extensive knowledge base \cite{chanSemanticallyFarInspirations2017}. However, real creativity often requires domain-specific and nuanced knowledge that might be absent in LLM training data. Hence, to ensure quality, LLM outputs for these tasks must be guided, selected, or edited by humans through precise human-AI interactions.

\subsection{Complex and Interdependent Tasks}

Beyond the atomic creative tasks are complex and interdependent tasks such as active search \cite{palaniCoNotateSuggestingQueries2021,palaniActiveSearchHypothesis2021}, neurocognitive disorder test \cite{dingTalkTiveConversationalAgent2022}, collaborative design work \cite{macneilFreeformTemplatesCombining2023}, and storytelling in text \cite{yuanWordcraftStoryWriting2022a, singhWhereHideStolen2022} and images \cite{yanXCreationGraphbasedCrossmodal2023}. These tasks demand not only domain expertise but also capabilities for planning, reasoning, ideating, and maintaining context over extended periods to create coherent and innovative content. Researchers have developed tools like Promptify \cite{bradePromptifyTexttoImageGeneration2023} and PromptPaint \cite{chungPromptPaintSteeringTexttoImage2023} for text-to-image generation, Spellburst \cite{angertSpellburstNodebasedInterface2023} for exploratory creative coding, and XCreation \cite{yanXCreationGraphbasedCrossmodal2023} for cross-modal storytelling to facilitate these intricate interactions.

\section{Preliminary Studies and Results}

Our research comprises three studies, each examining human-AI interactions across tasks varying in creativity and complexity. The initial study delves into fixed-scope content curation tasks, utilizing news headline generation as the investigative lens. The first study examines fixed-scope content curation tasks, employing news headline generation as the medium for exploration. The second study transitions to probing atomic creative tasks, with a focus on analogy generation to uncover underlying creative processes. Lastly, the third study escalates in complexity, exploring the complex and interdependent tasks associated with data visualization. These three studies collectively offer a comprehensive understanding of human-AI interaction across a spectrum of task creativity and complexity.

\subsection{Study 1 on Fixed-Scope Content Curation Tasks: News Headline Generation}

\begin{figure}
\centering
\includegraphics[width=0.8\linewidth]{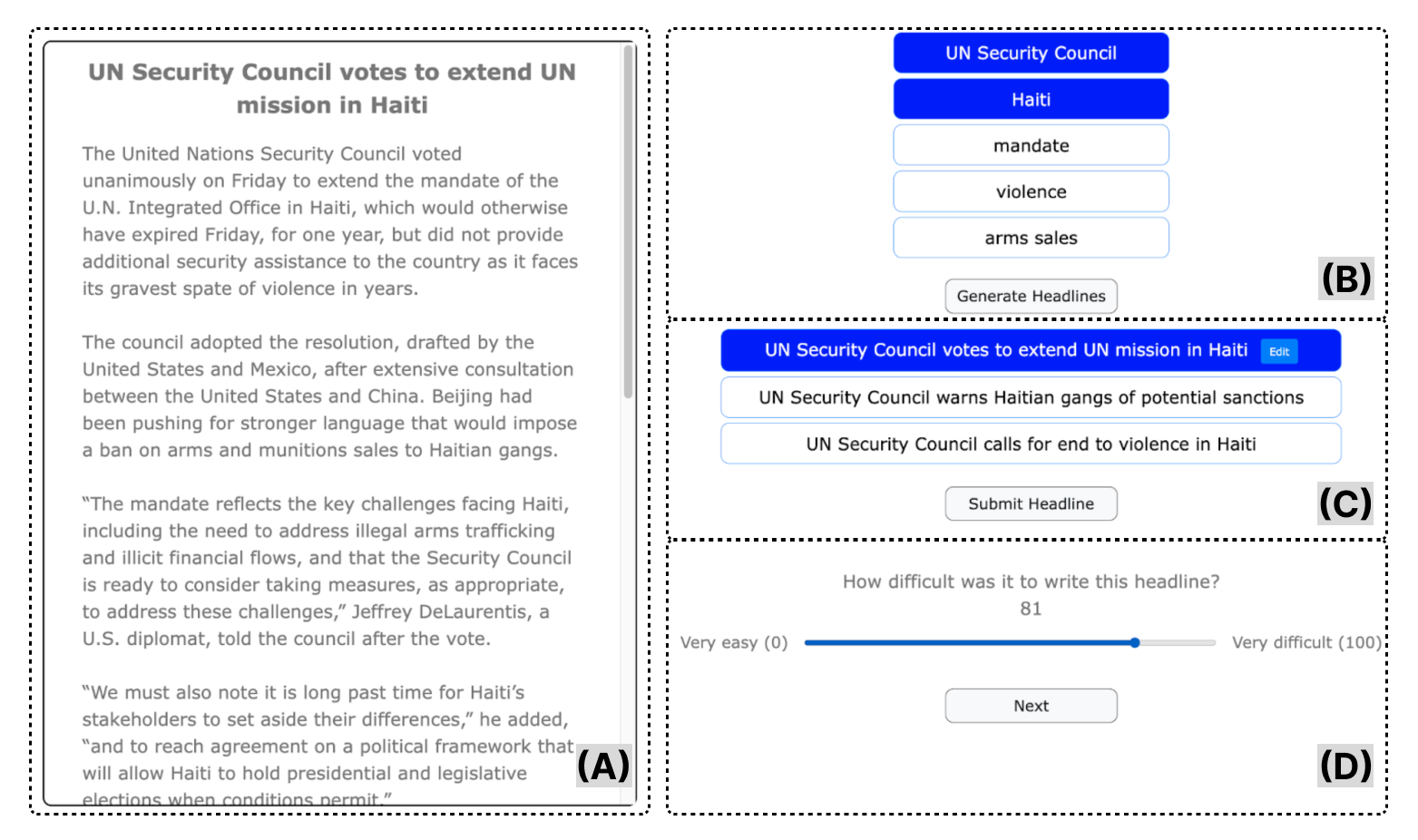}
\caption{Study 1 on fixed-scope content curation tasks - interface for human-AI news headline co-creation for \textit{guidance + selection + post-editing} condition: (A) news reading panel, (B) perspectives (keywords) selection panel (multiple keywords can be selected), (C) headline selection panel with post-editing capability, and (D) difficulty rating slider.}
\label{fig:study1_interface}
\end{figure}

This research investigates human-AI interactions within fixed-scope content curation tasks, specifically focusing on the generation of news headlines \cite{dingHarnessingPowerLLMs2023}. A between-subjects experiment was conducted involving 40 participants tasked with creating news headlines using either traditional manual methods or one of three LLM-enhanced systems: (1) \textit{selection}, (2) \textit{guidance + selection}, and (3) \textit{guidance + selection + post-editing}.
\begin{itemize}
    \item \textbf{Selection}: The LLM generates three headlines for each news article (\textit{generate headlines}), and the user selects the most appropriate one;
    \item \textbf{Guidance + Selection}: The LLM extracts several potential perspectives (keywords) from each news article (\textit{extract perspectives}), the user chooses one or more perspectives to emphasize in the headline, the LLM then generates three headlines for each news article based on the selected perspectives (\textit{generate headlines w/ perspectives}), and finally, the user selects the best one;
    \item \textbf{Guidance + Selection + Post-editing}: This is similar to \textit{Guidance + Selection}, but the user can further edit the selected headline (post-editing).
\end{itemize}
An example of the interface supporting \textit{guidance + selection + post-editing} is presented in Figure \ref{fig:study1_interface}.
Subsequently, the outputs produced via the five interaction types—manual, AI-only, and the three AI-assisted conditions—were evaluated by a cohort of 20 experts.

The results of the study indicate that LLMs (GPT-3 \textit{text-davinci-002} model), on average, are capable of generating headlines of a high quality independently. However, they are not without flaws, necessitating human oversight to correct inaccuracies. The study found that the \textit{guidance + selection} modality was particularly effective, facilitating the efficient generation of superior headlines compared to more labor-intensive methods such as extensive post-editing or entirely manual composition. Interestingly, the research also observed that across all conditions, there was a similar level of perceived trust and control reported by participants.

\subsection{Study 2 on Atomic Creative Tasks: Cross-Domain Analogy Generation}

\begin{figure}
\centering
\includegraphics[width=0.8\linewidth]{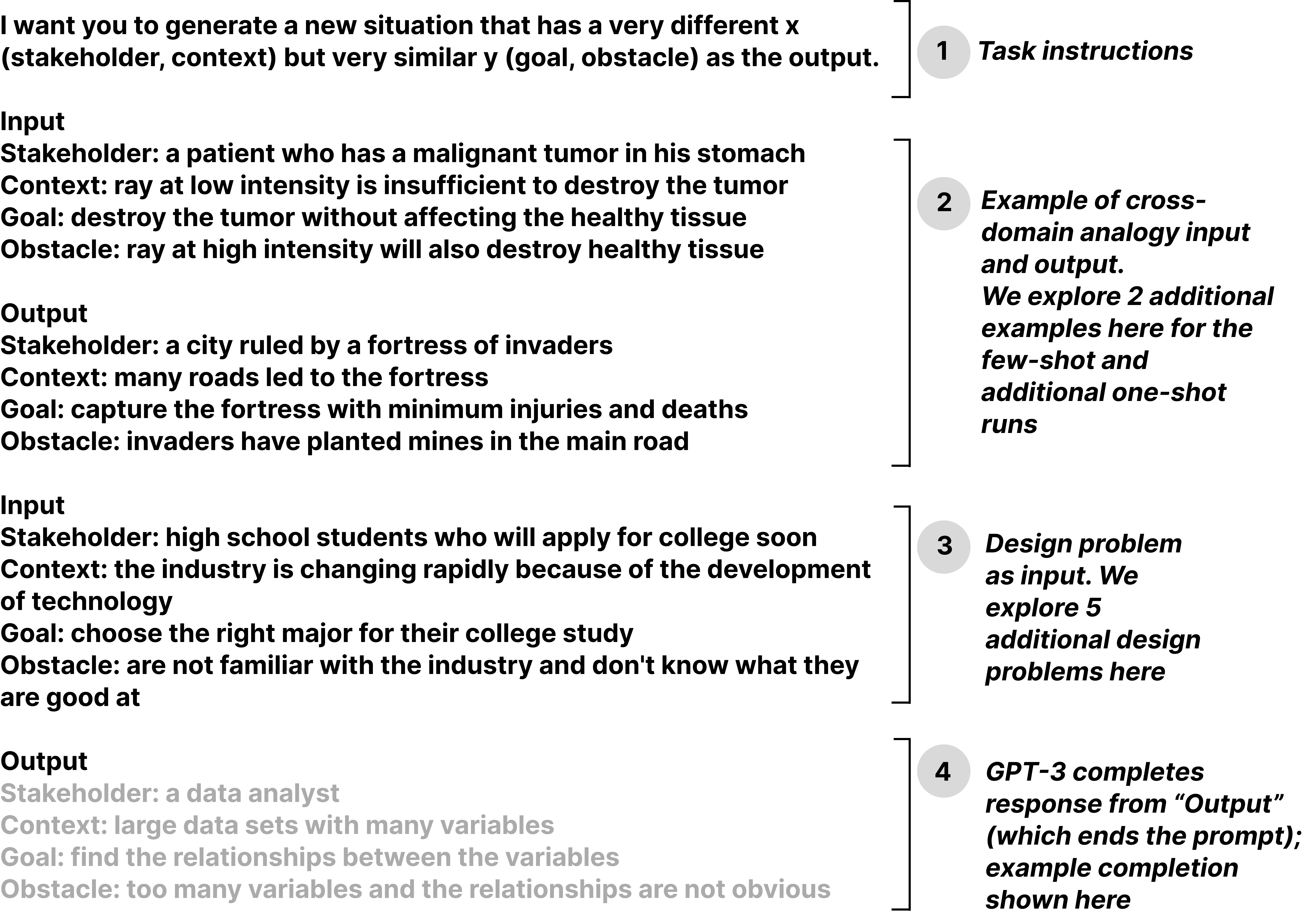}
\caption{Study 2 on atomic creative tasks - our learning prompts based on Duncker and Lees' radiation problem \cite{dunckerProblemsolving1945} for generating cross-domain analogies.}
\label{fig:prompt_structure}
\end{figure}

Transitioning from fixed-scope content curation tasks, the second study delves into atomic creative tasks, focusing on cross-domain analogy generation. In this context, classical analogous problems curated or crafted by humans, such as Duncker and Lees' radiation problem \cite{dunckerProblemsolving1945}, are employed to \textit{guide} Large Language Models (LLMs) as examples in generating analogies
(see Figure \ref{fig:prompt_structure})
. Our study indicates the necessity for human evaluators to \textit{select}, \textit{rate}, and \textit{post-edit} these AI-generated analogies to address potential biases and ensure ethical and legal compliance.

In our recent exploratory experiments assessing LLMs' efficacy in atomic creative tasks, we tasked participants with reformulating the original problem using 120 analogous problems generated by GPT-3 \textit{text-davinci-002} model \cite{ding2023fluid}. The results were encouraging: the AI-generated analogies were predominantly perceived as useful, with a median helpfulness rating of 4 out of 5, and they instigated substantial changes in problem formulation in about 80\% of cases. However, up to 25\% of the outputs were rated as potentially harmful, chiefly due to unsettling content that did not necessarily fall under bias or toxicity. These insights highlight the potential of LLMs to enhance atomic creative tasks and simultaneously draw attention to the necessity for human intervention in the form of selection, rating, and post-editing.

\subsection{Study 3 on Complex and Interdependent Tasks: Data Visualization}

\begin{figure*}
\includegraphics[width=0.8\textwidth]{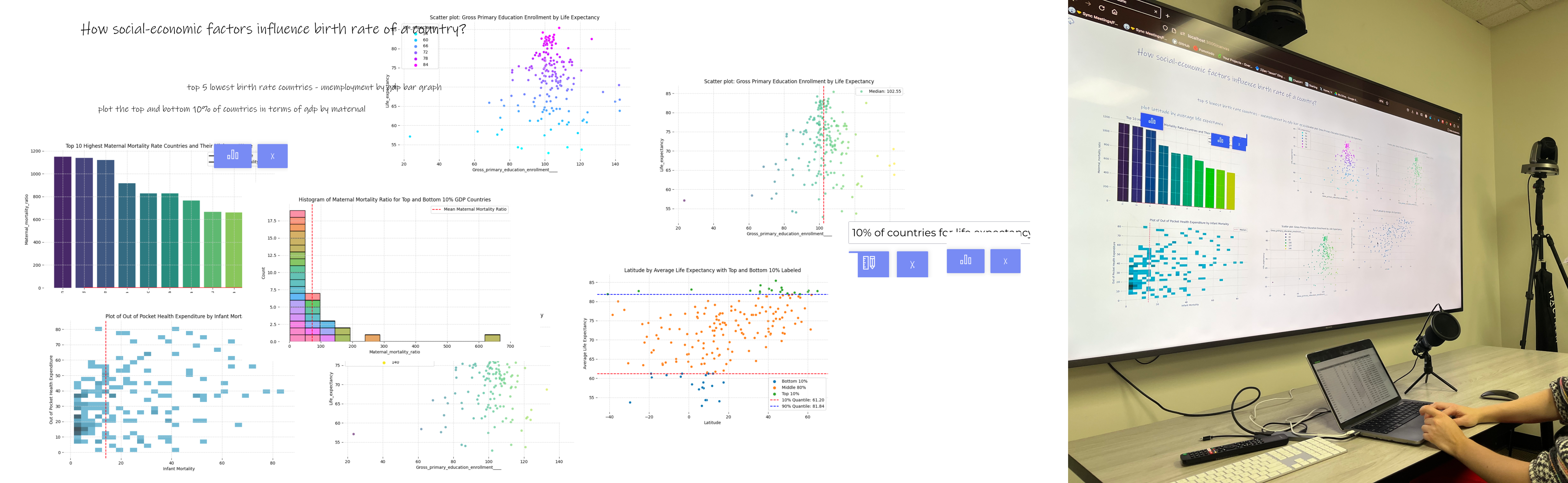}
  \caption{Study 3 on complex and interdependent tasks - data analysis results in design-like canvas environment: results (left) and study setting (right).
  }
  \Description{Study 3 - data analysis results in design-like canvas environment.}
  \label{fig:canvas}
\end{figure*}

This study aims to explore human-AI interactions within the context of complex and interdependent tasks, with an emphasis on data visualization. Exploratory data analysis typically involves searching for insights that extend beyond linear and sequential thinking. While tools like Jupyter Notebook offer substantial analytical capabilities, they often do not fully support the exploratory nature required for complex data analysis and corresponding human-AI interactions. To address these limitations, we introduced a "design-like" intelligent canvas. This tool integrates generative AI into the data analysis process and enhances human-AI collaboration by enabling interactive editing, which facilitates rapid prototyping, iteration, and the efficient management of comparative visualizations
(refer to Figure \ref{fig:canvas}). A user study involving 10 participants with in-depth interview was conducted to evaluate the interface's facilitation of human-AI interaction in tackling complex and interdependent tasks.

\section{Implications and future work}

Our upcoming research aims to explore further into the intricacies of human-AI interactions, particularly focusing on complex and interdependent tasks that exhibit a high potential for interaction. This research builds upon the findings from Study 3, which investigated the preliminary stages of data visualization prototyping facilitated by Generative AI. These initial attempts often led to an unstructured, "messy" output, as demonstrated by the outcomes of Study 3. Moreover, our objective extends beyond achieving a neatly organized set of data visualizations. Data analysis aims to uncover new insights, such as explanations for discrepancies observed when comparing visualizations. While the interpretation of these visualizations traditionally falls within the expertise of human data analysts, the advent of generative AI and its multi-modal understanding capabilities, including image captioning, presents an opportunity for AI to aid in interpreting and structuring visualizations. Thus, our research question is: \textbf{How can we harness multi-modal understanding of generative AI for real-time data visualization management?} To answer this question, we plan to conduct a formative study focusing on how experienced data analysts interpret and organize data visualizations as Knudsen et al. \cite{knudsenExploratoryStudyHow2012}. This formative study will be followed by the development of a mixed-initiative user interface designed for real-time data visualization management, and a user study to evaluate the performance of the system.

\bibliographystyle{ACM-Reference-Format}
\bibliography{sample-base}

\end{document}